\documentclass[aps,prl,twocolumn,showpacs,superscriptaddress,nobibnotes]{revtex4-2}

\usepackage{graphicx}
\usepackage[utf8]{inputenc}
\usepackage{siunitx}
\usepackage{amsmath}
\usepackage{amssymb}
\usepackage{enumerate}
\usepackage{color}
\usepackage{hyperref}
\usepackage[american]{babel}
\usepackage{etoolbox}
\usepackage{floatrow}
\usepackage{lineno,hyperref}
\usepackage{subfigure}
\usepackage[disable]{todonotes}
\usepackage{tikz}
\usepackage{nicefrac}

\patchcmd{\thebibliography}{\section*{\refname}}{}{}{}

\newcommand{\aegis}{AE$\bar{\hbox{g}}$IS}

\begin{document} 


\title{\LARGE\bf A fiber detector to monitor ortho-Ps formation and decay}

\newcommand{\corresponding}[1]{\altaffiliation{Corresponding author, #1}}

\newcommand{\afftn}[0]{\affiliation{Department of Physics, University of Trento, via Sommarive 14, 38123~Povo, Trento, Italy}}

\newcommand{\affinfntn}[0]{\affiliation{TIFPA/INFN Trento, via Sommarive 14, 38123~Povo, Trento, Italy}}

\newcommand{\affcern}[0]{\affiliation{Physics Department, CERN, 1211~Geneva~23, Switzerland}}

\newcommand{\affinfnpd}[0]{\affiliation{INFN Padova, via Marzolo~8, 35131~Padova, Italy}}

\newcommand{\affinfnpdL}[0]{\affiliation{Department of Physics and Astronomy ''Galileo Galilei'', University of Padova, Via Marzolo, 8, 35131~Padova, Italy}}

\newcommand{\affoslo}[0]{\affiliation{Gravitation Astroparticle Physics Amsterdam (GRAPPA), University of Amsterdam, Science Park 904, 1098~XH~Amsterdam, The~Netherlands}}

\author{B.~Rien\"{a}cker}
\corresponding{b.rienaecker@cern.ch}
\affcern

\author{R.~S.~Brusa}
\afftn
\affinfntn

\author{R.~Caravita}
\affinfntn



\author{S.~Mariazzi}
\affinfntn

\author{L.~Penasa}
\afftn
\affinfntn

\author{F.~Pino}
\affinfnpdL

\author{O.~A.~Ranum}
\affoslo

\author{G.~Nebbia}
\affinfnpd

\noaffiliation{}

\date{\today}
      
\begin{abstract}
We describe a novel method to use a scintillating fiber detector similar to the Fast Annihilation Cryogenic Tracking (FACT) used at the antimatter experiment \aegis{} to monitor the presence of ortho-positronium. A single scintillating fiber was coupled to a photomultiplier tube and irradiated by flashes of about \SI{6e6}{} \SI{511}{\kilo\electronvolt} $\gamma$-rays produced by $\approx$\SI{10}{\nano\second} long positron pulses.
The results were used to demonstrate the ability to track the creation and annihilation of ortho-positronium atoms over time in cryogenic and highly magnetic environments by using the FACT detector as a “digital calorimeter”.
\end{abstract}

\pacs{36.10.Dr, 78.70.Bj, 29.40.Mc}
\maketitle{}

\section{Introduction}

Positronium (Ps), the bound state of an electron and a positron, is a purely leptonic matter-antimatter system with many applications in fundamental research \cite{Mil01, Ito2005:Ps_porosimetry, Cass07:Ps2, cassidy_review:18, Mills2019:PsBEC, Nagashima2020, Zimmer2021:PsCooling}. In particular, positronium can be used to induce the charge exchange reaction with cold trapped antiprotons in order to efficiently create antihydrogen. This was recently achieved by \aegis{} (Antimatter Experiment: Gravity, Interferometry, Spectroscopy) \cite{AEGIS2021:HBAR} and is also planned to be used by GBAR (Gravitational Behaviour of Antihydrogen at Rest) \cite{GBAR:Ps_Hbar}. Both experiments are located at the Antiproton Decelerator (AD) facility at CERN. The \aegis{} collaboration produces ground state positronium by implanting positron bunches into a nanochanneled silicon target \cite{mariazzi_prl:10, AEGIS2021:Morpho}. A substantial fraction of the long-lived positronium atoms in a triplet state (ortho-positronium with total spin $S=1$) subsequently cools by inelastic collisions with the channel walls and escapes from the nanoporous target at almost environmental temperature into vacuum. Such cold Ps is then excited to Rydberg states by laser irradiation \cite{aegis_neq3:16} before reaching the trapped antiproton plasma and forming antihydrogen. Such experiments require ultra-high vacuum, a homogeneous magnetic field in the order of few Tesla and cryogenic temperatures, making it difficult to set up an efficient detector to monitor and track positronium creation and decay.

The commonly used technique for Ps detection, namely Single Shot Positron Annihilaton Lifetime Spectroscopy (SSPALS) with scintillation detectors \cite{CassidySSPALS2007}, is stringent limited in narrow-spaced environments at liquid helium temperatures with high magnetic fields and small coverable solid angles such as is occurring at \aegis{} \cite{aegis_SSPALS:2019,aegis_MCP:2020}.
Therefore, positronium formation is monitored in a destructive way by laser-induced photoionization and by imaging the ionized positrons, which are axially confined by a \SI{1}{\tesla} magnetic field, with a microchannel plate and a phosphor screen \cite{aegis_MCP:2019,AEGIS:2020_RydbergPs}. Antihydrogen formation and annihilation, on the other hand, can be monitored by means of the Fast Annihilation Cryogenic Tracking (FACT) scintillating fiber detector surrounding the production region \cite{aegis_fact:13,aegis_FACT:2020}. The FACT detector has been designed to detect pions produced by the annihilation of antihydrogen atoms. Nevertheless, the scintillating fibers could be used as well to non-destructively monitor the production and the excitation of ortho-positronium by the detection of $3\gamma$ annihilation following the in-flight decay of o-Ps in vacuum.
In order to determine such a possibility, we first ran Monte Carlo simulations of the response of such a fiber detector to gamma radiation. We then performed experimental tests on one of the scintillating fibers as is installed in the FACT detector and studied its response to positron bursts from the \aegis{} positron system.

\section{A "digital calorimeter" for o-Ps: Principle and simulations}

\begin{figure}[hptb]
    \centering
    \includegraphics[width=0.85\linewidth]{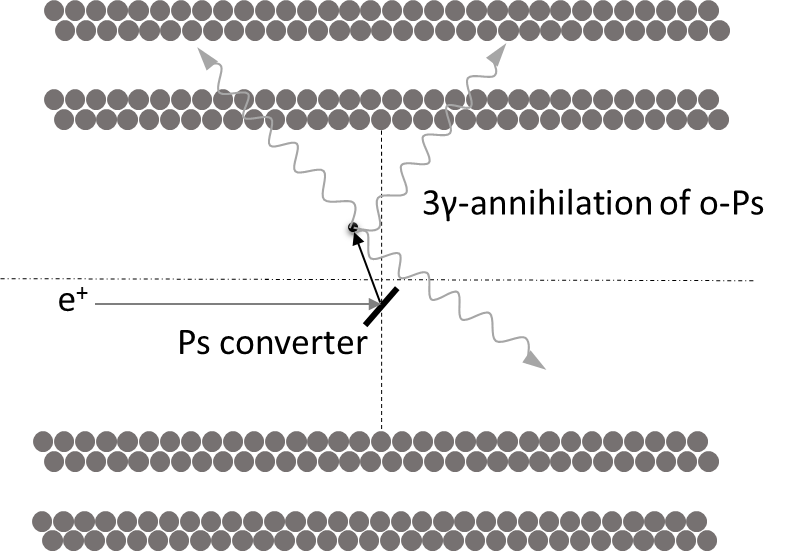}
    \caption{Schematic of the FACT cross section. In total 794 scintillating fibers in four layers in cylindrical geometry (inner diameter is \SI{70}{\milli\meter}, outer diameter is \SI{98}{\milli\meter}) are used to detect annihilation products. The Ps converter is positioned in the center, but radially displaced by \SI{20}{\milli\meter} to be aligned with \aegis{} positron transfer line.}
    \label{fig:FACTgeo}
\end{figure}

The FACT detector of \aegis{} is made of 794 scintillating fibers distributed in four layers with cylindrical symmetry and a length of \SI{200}{\milli\meter}. The fiber type is \emph{Kuraray SCSF-78M}, multi-cladded with \SI{1}{\milli\meter} in diameter. There are two layers of scintillating fibers at radial distances of \SI{70}{\milli\meter} and \SI{98}{\milli\meter} from the central beam axis. Each layer consists again of two layers, but one is shifted against the other in order to avoid blind spots in the detection area. Per single layer, the fiber centers are horizontally separated by \SI{1.2}{\milli\meter} and the radial distance of the shifted layer is increased by \SI{0.8}{\milli\meter} (see Fig. \ref{fig:FACTgeo}). 

Each scintillating fiber is mechanically and optically coupled to a clear fiber, which transports scintillation light to a Multi-Pixel Photon Counter (MPPC, type \emph{Hamamatsu S10362-11-100C}), which produces a small electrical signal at its output. A hit is registered only when its signal overcomes a given threshold. The MPPC output is connected to a fast monolithic amplifier (\emph{Mini-Circuits MAR-6+}), whose signal is fed into a discriminator which then returns the time over threshold (ToT), digitized by a FPGA acquisition board \cite{aegis_fact:13,aegis_fact:20}. 
The positron/Ps converter is positioned in the center of FACT, but the radial distance from the central axis is approximately \SI{20}{\milli\meter} so that it is aligned with the positron transfer line. When positrons hit the converter, ortho-positronium is created which flies for several millimeters before it annihilates into predominately three gamma quanta with energies ranging between 0 and \SI{511}{\kilo\electronvolt}.
This continuous o-Ps annihilation spectrum is described by the so-called Ore-Powell formula \cite{Ore49}, which was used for the following Monte Carlo simulation.

\begin{figure}[hptb]
    \centering
    \includegraphics[width=1\linewidth]{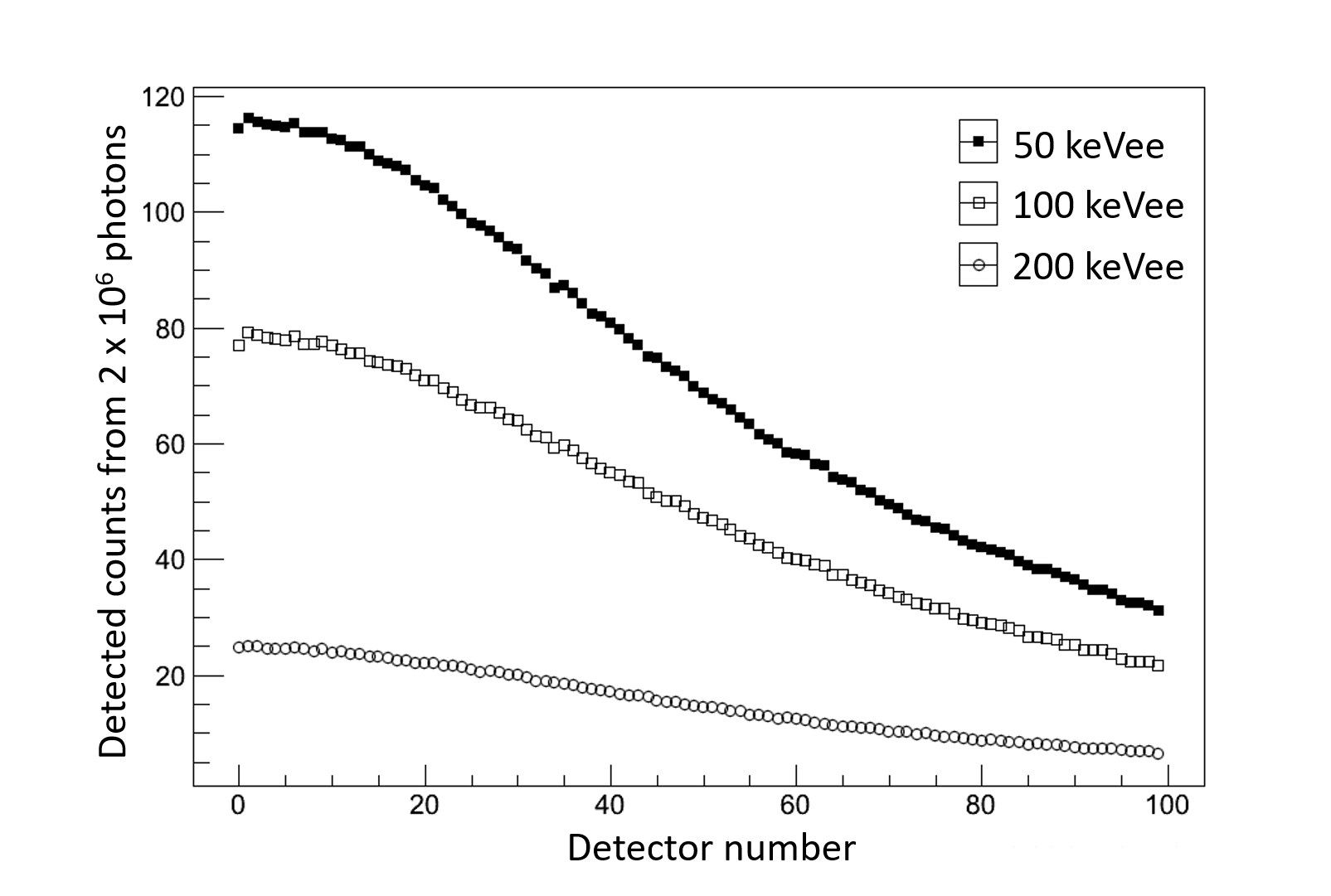}
    \caption{Simulated response of an array of scintillating fibers to a pulse of \SI{2e6}{} o-Ps annihilation gamma quanta with applied energy thresholds of 50, 100 and 200\,keVee, respectively.}
    \label{fig:PENELOPE}
\end{figure}

With the 2014 version of the PENELOPE code \cite{PENELOPE} we simulated the response of an array of scintillating fibers in the same geometry as FACT to a discrete Ore-Powell energy spectrum of $\gamma$-radiation. Fig. \ref{fig:PENELOPE} shows the simulated detector response to a delta pulse of \SI{2e6}{} gamma quanta originating from the target region for three different energy thresholds. The vertical axis represents the number of gamma quanta leaving a signal above the threshold in each fiber. Exploiting the symmetry of the entire setup, only 1/8 of all available fibers needed to be simulated for this proof-of-principle. They were numbered from 0 to 99, with the central fiber of the entire array being placed at the zero position of the abscissa in the simulation. 
The three different curves show the number of gamma quanta leaving more than \SI{50}{\kilo\electronvolt ee} (black squares), \SI{100}{\kilo\electronvolt ee} (white squares) or \SI{200}{\kilo\electronvolt ee} (white circles) of energy in each fiber. The reduction of the number of counts reflects the effect of the solid angle, which is decreasing by approximately \SI{70}{\percent} when going from the central fiber (\#0) to the outermost fiber (\#99). This basic response signal is valid for one layer of fibers with constant radius. 
 
Considering its solid angle $\Omega \approx \SI{0.09}{sr}$, the central fiber would be irradiated by \SI{1.4e4}{} photons originating from the position of the Ps converter. However, the simulation result with an energy threshold of \SI{50}{\kilo\electronvolt ee} as plotted in Fig. \ref{fig:PENELOPE} shows that only 115 gamma quanta leave a sizeable signal inside the fiber. This is mainly due to the intrinsic efficiency of the fiber yielding the ability to "switch it on", but of course also partly due to the applied threshold.
In order to validate the simulation, we estimated this intrinsic efficiency of one fiber at a given energy threshold by taking into account the solid angle of the central fiber, and compared this to measurements existing in literature for similar systems. The intrinsic efficiency of one fiber at an energy threshold of \SI{150}{\kilo\electronvolt ee} is \SI{0.3}{\percent} as yielded by the PENELOPE simulation. This compares well with the efficiency of \SI{0.23}{\percent} measured in Ref. \cite{Machaj2011} on a \SI{1}{\milli\meter} thick plastic scintillator for gamma radiation of the same energy regime and detection threshold.

\begin{figure}[hptb]
    \centering
    \includegraphics[width=1\linewidth]{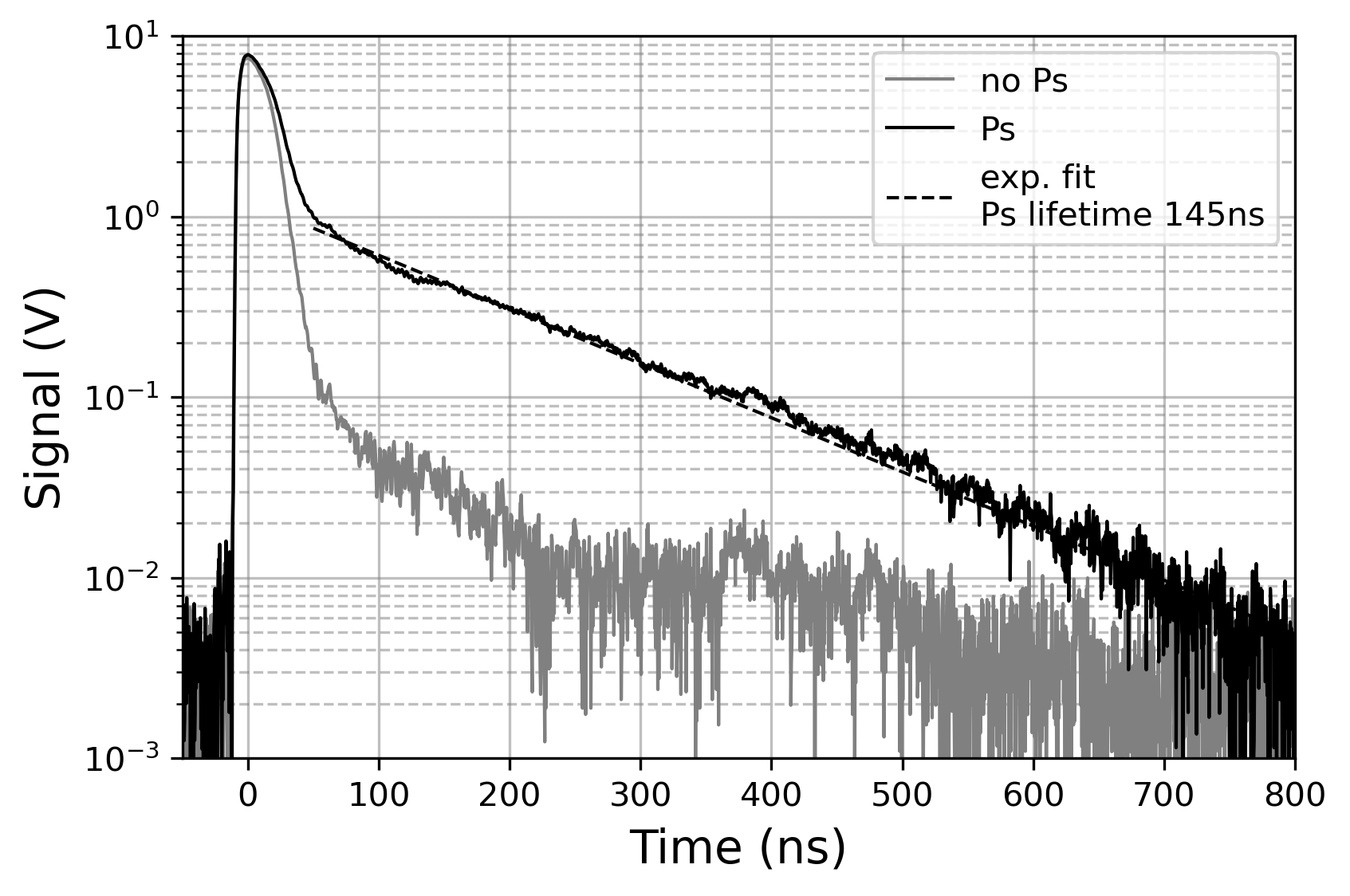}
    \caption{Examples for SSPALS spectra measured with a PbWO$_4$ scintillation detector for an aluminum  target (no Ps formation, grey curve) and for a positron/Ps converter (Ps formation, black curve). The SSPALS spectrum in presence of the converter shows an exponentially decreasing Ps tail with the typical ground state vacuum lifetime of about \SI{142}{\nano\second}. The best exponential fit is reported as the dashed line. The slight deviation from the theoretical Ps lifetime value comes from systematic effects such as the non-linear detector response function at high signal amplitudes.}
    \label{fig:SSPALS}
\end{figure}

In order to be able to simulate the detector response to a decaying cloud of Ps atoms, we first have to consider a typical ortho-positronium SSPALS spectrum as is shown in Fig. \ref{fig:SSPALS}. This spectrum was obtained by implanting a pulse of \SI{3e6}{} positrons with a time spread of about \SI{10}{\nano\second} full-width-half-maximum (FWHM) and with a kinetic energy of roughly \SI{3}{\kilo\electronvolt} into a positron/Ps converter target. 
The annihilation signal was recorded with a PbWO$_4$ scintillation detector positioned about \SI{4}{\centi\meter} away from the Ps converter target inside a dedicated test chamber. The used positron system and the Ps converter are described in detail elsewhere \cite{aegis_nimb:15,AEGIS2021:Morpho}. 
As shown in Fig. \ref{fig:SSPALS}, the positrons hit the Ps converter target at time $t = 0$, leading to an almost instantaneous peak of positron annihilation into \SI{511}{\kilo\electronvolt} gamma quanta. In the absence of o-Ps formation, the signal quickly approaches the noise level. Conversely, in the presence of o-Ps being emitted into vacuum, the signal at times greater than \SI{50}{\nano\second} is proportional to the number of gamma quanta emitted by the decaying cloud of o-Ps as a function of time with the characteristic ground state vacuum lifetime of \SI{142}{\nano\second}. Due to the exponential nature of the decay, one can calculate the number of annihilation products occurring in any time interval of a chosen width as a function of the initial number of ortho-Ps atoms.

We then take into account the \SI{200}{\mega\hertz} clock as used in the FACT acquisition system, which sets a lower limit for the sampling time of \SI{5}{\nano\second} \cite{aegis_fact:13}. We measured the average recovery time $\tau_{\text{MPPC}}$ of one Hamamatsu MPPC and its amplifier with an oscilloscope, amounting to less than \SI{5}{\nano\second} following the detection of a single photon. We thus chose a time integration interval of \SI{20}{\nano\second} for the simulation of the o-Ps decay, which surely is a long enough recovery time for the readout electronics after detecting single photons. In case too many photons arrived at the fiber within this chosen time window, one would adjust the threshold on the corresponding MPPC until the recovery time is again within an acceptable range. 
If we now count the number of fibers firing above that threshold inside the chosen time interval and track this activity over time, while taking into account the position of each fiber and thus the resulting solid angles, we will finally reproduce the exponentially decaying number of o-Ps atoms. 
In other words, the response to the o-Ps decay hinges on the statistical nature of fibers being switched \emph{on} or \emph{off} under the exponentially decreasing bombardment with annihilation gamma quanta. Note that the choice of an optimal threshold for each MPPC is crucial, as too high values cut away the important long tail of o-Ps decay, while a too low threshold would saturate the MPPCs for a too long time --- up to several \SI{100}{\nano\second} --- due to the intense prompt positron $2\gamma$-annihilation pulse.
If there was no formation of Ps in the target, all implanted positrons would rapidly annihilate into $2\gamma$ quanta, producing a spectrum that shows a large number of fibers firing  at $t=0$, and then rapidly approaching noise level within the first 50-\SI{100}{\nano\second}. 

This noise level in turn can be estimated using the dark noise level of the real FACT detector at low energy thresholds, which amounts to \SI{100}{\kilo\hertz} per channel \cite{aegis_fact:13}. This means, each of the 794 fibers fires $0.002$ times within a time interval of \SI{20}{\nano\second}. The way this will affect the counting when taking into account all fibers is described by a Poisson distribution, with an expectation value $\lambda=794\cdot0.002=1.6$ counts within a \SI{20}{\nano\second} time interval as was used for the simulation. Thus if the recorded counts  exceed the noise level several tens of nanoseconds after positron implantation, it must be due to o-Ps annihilation.

In summary, the method proposed here consists of counting the number of fibers arranged in a cylindrical array that are firing within a fixed time interval, e.g. \SI{20}{\nano\second} as explained above, and tracking such a number as a function of the time. Taking into account the geometry of the detector array, one can find the exponentially decreasing number of gamma quanta originating from o-Ps annihilation in flight and even infer the approximate amount of originally formed positronium atoms.

\section{Experiments with a single fiber}

We coupled one Kuraray scintillating fiber to a fast photomultiplier tube (PMT \emph{Hamamatsu R1450}). The characteristics of such a fiber-PMT assembly are shown in Table \ref{tab:stats} on the left side. This should be compared to the characteristics of the MPPC on the right side as used in the FACT detector of \aegis{}. Both systems are very similar, therefore the results we obtain with our fiber-PMT assembly are indicative for the combination of fiber-MPPC as used in FACT.

\begin{table}[hptb]
    \centering
    \begin{tabular}{l|c|c}
         &  Fiber-PMT & Fiber-MPPC\\
         \hline
         Spectral response &  $300-\SI{600}{\nano\meter}$ &$320-\SI{900}{\nano\meter}$\\
         $\lambda$ of max. response & \SI{420}{\nano\meter} & \SI{440}{\nano\meter} \\
         Nominal voltage & \SI{-1500}{\volt} & \SI{70}{\volt} \\
         Gain & \SI{1.7e6}{} & \SI{2.4e6}{} \\
         Rise time & \SI{1.8}{\nano\second} & \SI{1.2}{\nano\second} \\
         Decay time & \SI{6}{\nano\second} & $<$\SI{5}{\nano\second} \\
         \hline
    \end{tabular}
    \caption{Characteristics of systems where a PMT and a MPPC are coupled to a scintillating fiber. Both systems are very similar, thus results obtained with fiber-PMT can directly be compared to fiber-MPPC as is used for the FACT detector.}
    \label{tab:stats}
\end{table}

The fiber was shielded from ambient light by a black plastic coverage. One of the polished ends was coupled to the PMT via a black plastic holder by an air-gap without use of optical grease or bonding. This fiber-PMT assembly was tested by using a $^{133}$Ba source whose emission spectrum shows a dominant peak at \SI{356}{\kilo\electronvolt} energy \cite{133Ba}. The X-ray peaks at around \SI{31}{\kilo\electronvolt} energy were strongly suppressed, since the source was sealed in a brass casing of about \SI{5}{\milli\meter} thickness. As a consequence, the dominant peak of the Ba-spectrum can be used as a proxy for the mean energy of the continuous annihilation spectrum of o-Ps. By recording the number of counts with a Tektronix TDS5054B oscilloscope, its trigger threshold being set to \SI{5}{\milli\volt}, we confirmed the ability of a \SI{1}{\milli\meter} thick Kuraray scintillating fiber to detect gamma rays in the energy range between $300-$\SI{400}{\kilo\electronvolt}. When the Ba source was located about \SI{10}{\milli\meter} away from the fiber, the background count rate of \SI{150}{\hertz} increased to several \SI{}{\kilo\hertz}. 

\begin{figure}[hptb]
    \centering
    \includegraphics[width=0.65\linewidth]{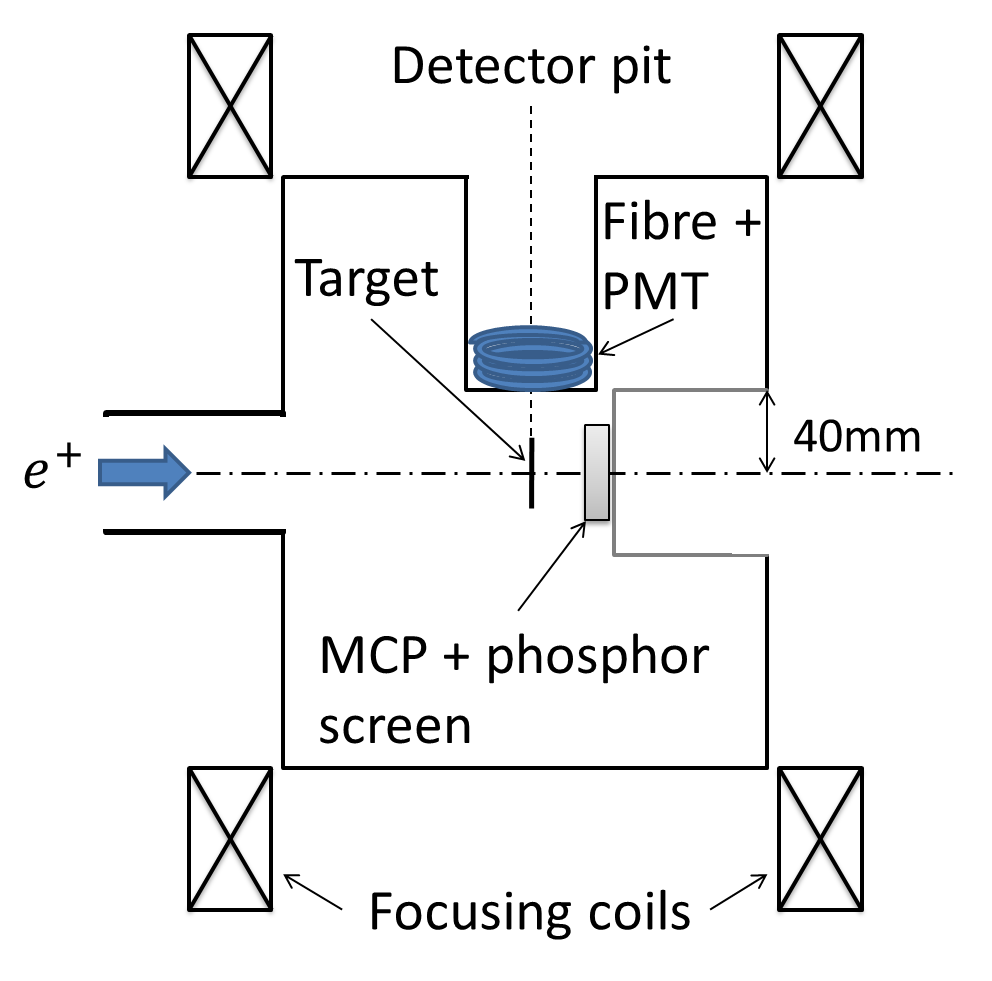}
    \caption{Test chamber, where the Kuraray scintillating fiber is tested with a pulse of \SI{3e6}{} positrons.}
    \label{fig:BB}
\end{figure}

In order to experimentally test the response of a Kuraray fiber to a burst of positrons, the aforementioned positron test chamber of \aegis{} (sketched in Fig. \ref{fig:BB}) was used. Pulses with about \SI{3e6}{} positrons were produced using the \aegis{} positron system, which is located at the Antiproton Decelerator (AD) facility at CERN. The loose end of the fiber was rolled up into three loops of about \SI{5}{\centi\meter} in diameter for a total length of about \SI{47}{\centi\meter}. The 3-loop end of the fiber has been inserted in the detector “pit” of the positron test chamber of \aegis{} with the loop axis pointing towards an aluminium target. The positron pulses were steered onto the target, which led on average to a peak amplitude of about \SI{100}{\milli\volt} on the used oscilloscope. This resembled the average response of our fiber-PMT assembly to the full $2\gamma$ annihilation signal consisting of about \SI{6e6}{} gamma quanta with \SI{511}{\kilo\electronvolt} energy. 
We assumed all positrons to annihilate on the surface of the aluminum target into two gamma quanta with no production of Ps or positron back-reflection. The background signal was acquired by switching off the magnets in the transfer line so that no positrons were transferred from the source region to the test chamber.

\begin{figure}[hptb]
    \centering
    \includegraphics[width=1\linewidth]{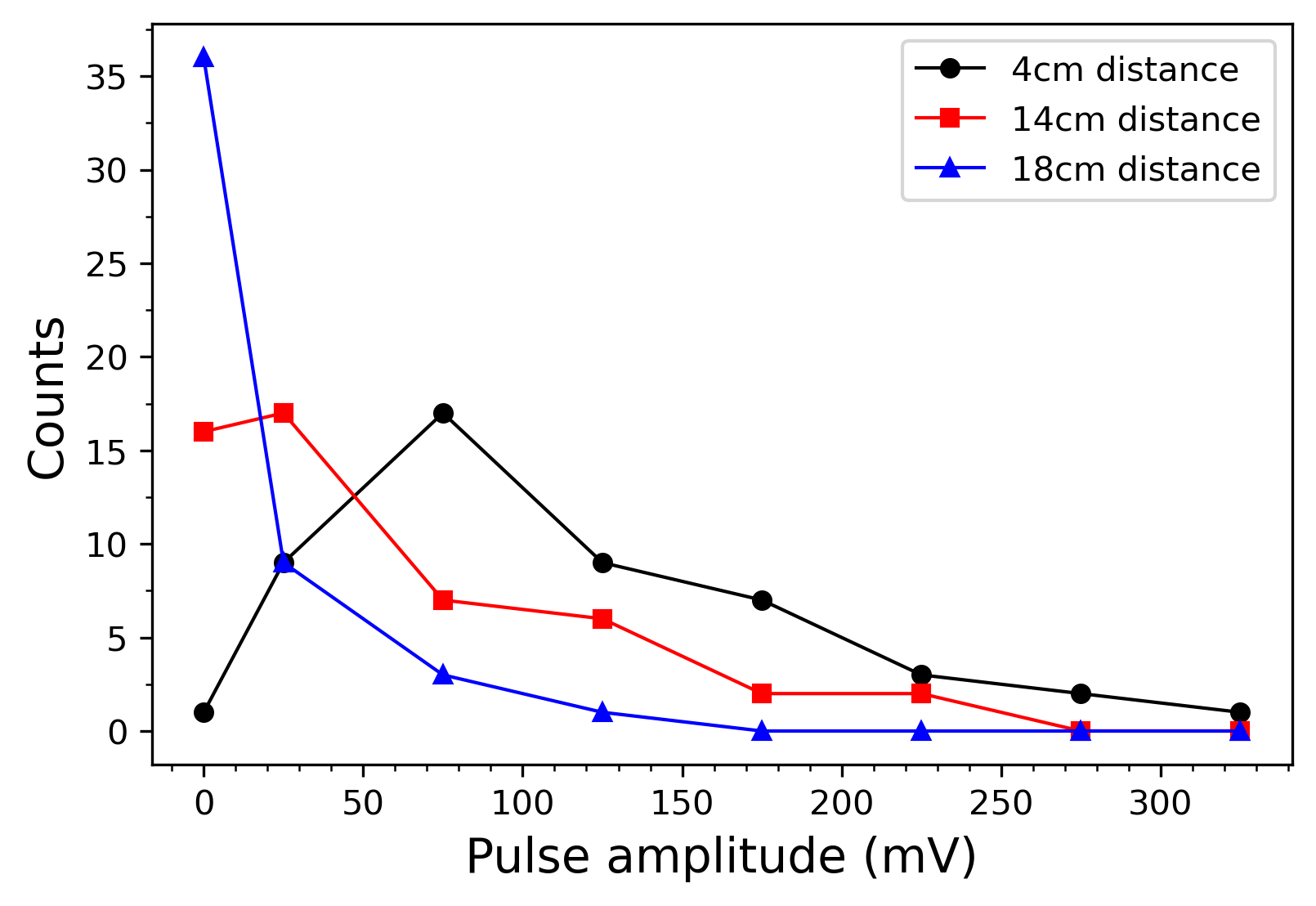}
    \caption{Number of occurrences as a function of pulse amplitudes for different distances, i.e. solid angles. Error bars are left out, as this is a qualitative indication only. Solid lines are meant as a guide to the eye.}
    \label{fig:distance}
\end{figure}

The 3-loop fiber was positioned in the detector pit at distances of 4, 14, \SI{18}{\centi\meter} from the aluminium target, and between 50 and 100 measurements were taken for each position recording the signal amplitudes. The resulting histograms are shown in Fig. \ref{fig:distance}. One can see that as distances from the target increase the distributions shift towards smaller signal amplitudes and have a smaller width. Moreover, the output of the fiber at greater distances was increasingly often not distinguishable from the noise level, which then added to the histogram as a zero signal. This was due to the decreasing solid angle covered by the 3-loop fiber in each position. The solid angles relative to the tested distances are \SI{0.161}{sr}, 
\SI{0.013}{sr} and \SI{0.008}{sr}, respectively. 

\begin{figure}[hptb]
    \centering
    \includegraphics[width=1\linewidth]{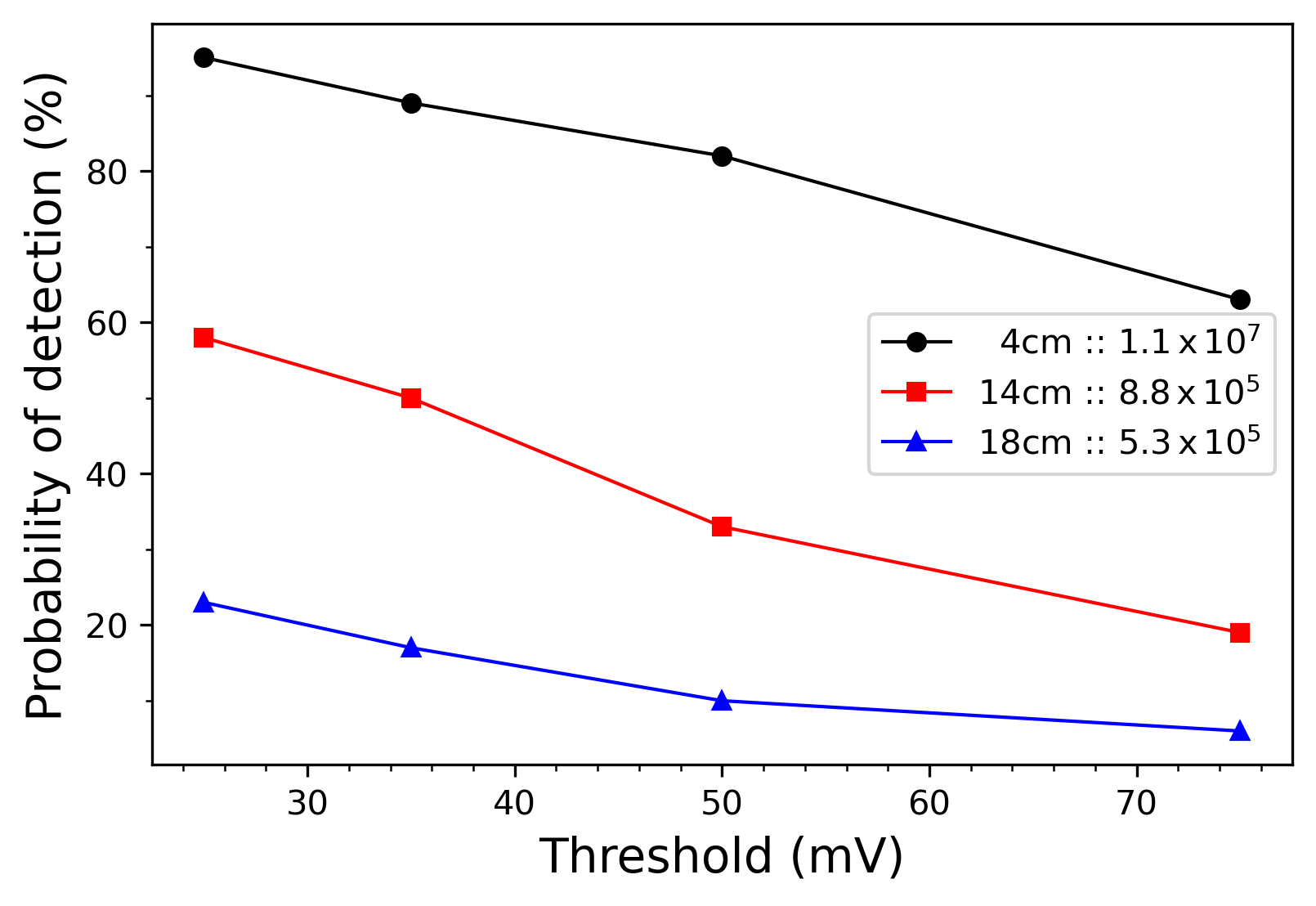}
    \caption{Probability of detecting a constant flash of \SI{6e6}{} photons from different distances as a function of the detection threshold. Equivalently, this can be interpreted as the probability to detect bursts of photons with varying intensities, but from a constant arbitrarily chosen solid angle (here \SI{0.09}{sr}). Again the error bars were left out for this qualitative result.}
    \label{fig:threshold}
\end{figure}

We can now derive the probability for a single fiber to register a signal above a certain energy threshold by calculating for each distance the ratio between the number of counts higher than a given threshold and the sum of all occurrences. In Fig. \ref{fig:threshold} we report the percent-probability that a fiber activates, i.e. that a photon burst with constant intensity deposits enough energy into the fiber that it exceeds the threshold. The corresponding thresholds are indicated in the abscissa.

\section{Simulated response of the "digital calorimeter"}

The variation of the distance and therefore of the solid angle can be thought to be equivalent to a change in the number of annihilating positrons. In other words, the measurements at different distances simulate the responses of one particular fiber with a fixed solid angle to flashes of gamma radiation with \emph{different} intensities --- exactly as it occurs during the exponential decay of o-Ps in free flight. Referring to the central fiber of a detector with the geometry of FACT, i.e. with a solid angle of \SI{0.09}{sr}, one can directly use these sets of measurements at different distances in order to obtain the probability of this fiber to respond to nearly instantaneous flashes of gamma quanta with varying intensities.
The curves in Fig. \ref{fig:threshold} can therefore also be interpreted as the percent-probability that a fiber covering a solid angle of \SI{0.09}{sr} activates due to a flash of \SI{1.1e7}{} (circles), \SI{8.8e5}{} (rectangles) or \SI{5.3e5}{} (triangles) annihilation photons, respectively.

We can now construct the response of the central fiber to a decaying cloud of o-Ps atoms at different times. The positronium cloud is generated by a pulse of $N_0$ positrons implanted at time $t=0$. It then exponentially decays with its typical vacuum annihilation lifetime of \SI{142}{\nano\second}. 
The initial o-Ps conversion ratio of the positron/Ps converter targets at \aegis{} is 0.3 \cite{AEGIS2021:Morpho}. Then taking only the fraction surviving magnetic quenching inside the present \SI{1}{\tesla} field (\SI{66}{\percent}), one finally finds a conversion efficiency of \SI{20}{\percent}.
From this, we obtained the number of gamma quanta emitted within intervals of \SI{20}{\nano\second} at any given time.
As we have shown above, a fiber has a chance to get activated with an certain intrinsic efficiency that depends on the set energy threshold when flashes of gamma rays irradiate it. Combining this effect with the simulated geometrical responses of the entire fiber array as was given for three thresholds in Fig. \ref{fig:PENELOPE}, we obtained the statistical number of scintillating fibers being activated at the same time, $T_{firing}$.

\begin{figure}[hptb]
    \centering
    \includegraphics[width=1\linewidth]{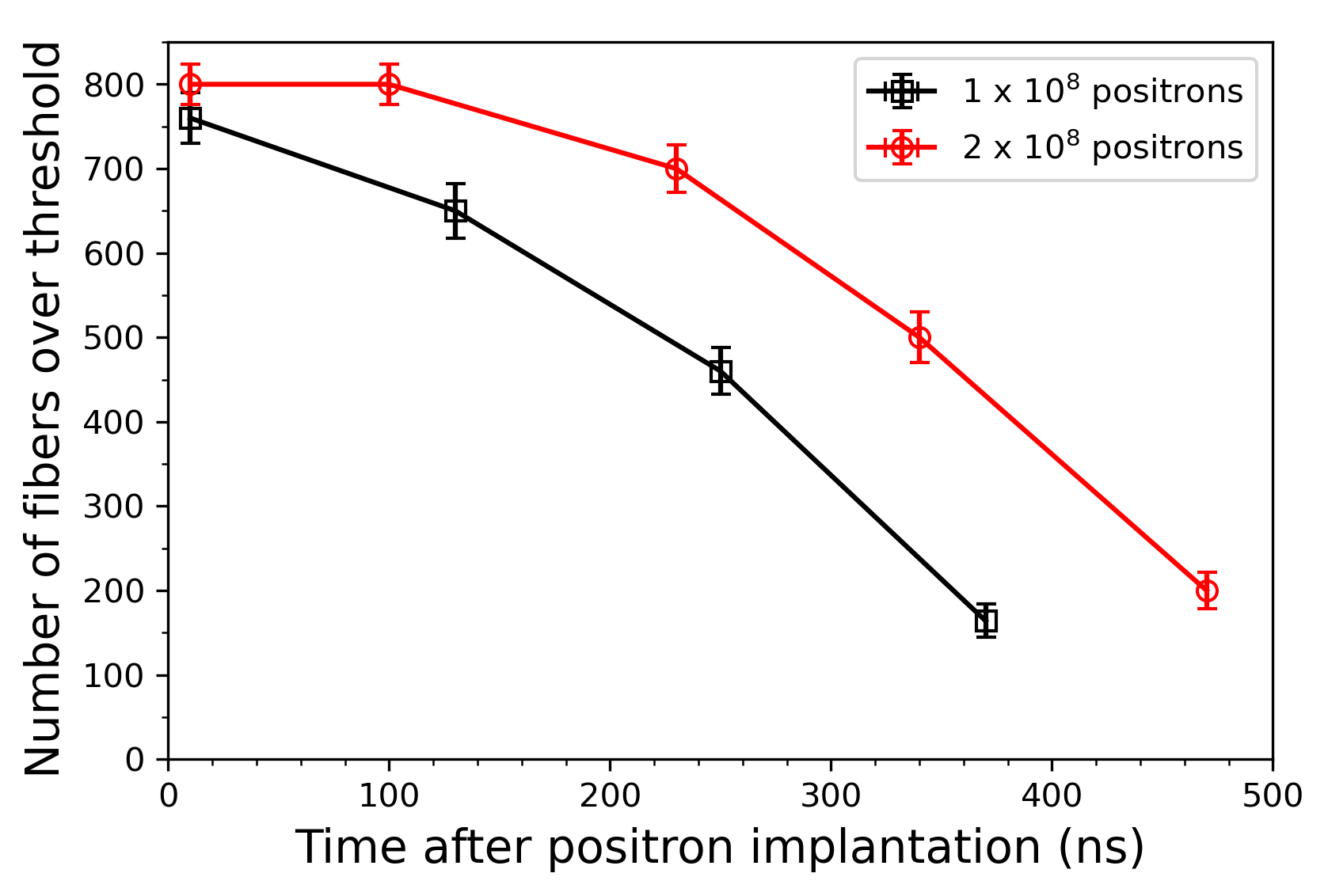}
    \caption{Simulated response of a fiber array detector such as FACT to \SI{1e8}{} positrons (circles) and \SI{2e8}{} positron (squares) as a function of time after positron implantation and conversion to Ps. As threshold \SI{25}{\milli\volt} was used for this test.}
    \label{fig:firing}
\end{figure}

The outcome is given in Fig. \ref{fig:firing}, which shows the number of fibers firing during the production and the decay of o-Ps for two different intensities of the initial positrons pulse at $t=0$. With the threshold set to \SI{25}{\milli\volt} and a positron pulse containing \SI{2e8}{} particles, the number of active fibers saturates for about \SI{100}{\nano\second} and then it decreases. When using only half the positrons, we observe no saturation but the decreasing tail. The threshold of \SI{25}{\milli\volt} has been chosen for this specific example and was used to verify the correct behaviour of the simulation. In reality, it depends on the intrinsic response of the single fiber-amplifier assembly, the number of implanted positrons and the aspired geometry and thus needs to be adjusted when using another system as is presented here. Notably, by regarding the convolution with the detector geometry, the resulting decrease of active fibers resembles the decay of o-Ps in free flight such as is usually tracked by the aforementioned SSPALS method (black curve in Fig. \ref{fig:SSPALS}). 
An estimate of the errors has been done for these particular tests by taking into account the statistical fluctuation on the number of gamma rays that leave a sizeable amount of energy, i.e. above the chosen \SI{25}{\milli\volt} threshold, in the central fiber. This yielded an error ranging from $3-$\SI{4}{\percent} at short times where almost all the fibers are hit and respond, to about $10-$\SI{12}{\percent} at the longest times where a reduced number of fibers respond. 

\section{CONCLUSIONS}

A novel method to use an array of scintillating fibers coupled to an amplification and detection system in order to track ortho-Ps formation and decay in cryogenic and magnetic environments over time has been tested. A single fiber coupled to a fast PMT was irradiated by flashes of about \SI{6e6}{} gamma quanta produced by a burst of positrons from the \aegis{} positron system. 
The response of the fiber obtained at different distances from the radiation source demonstrated the possibility to setup a fiber array in the geometry of the FACT detector of \aegis{} and use it to monitor the annihilation signal of decaying o-Ps in free flight. 
By introducing the precise geometry and efficiencies of a cylindrical fiber array as is used in FACT into a Monte Carlo simulation, we provided a proof-of-principle on the capability of this scintillating fiber detector to monitor the o-Ps formation and decay with a time resolution of at least \SI{20}{\nano\second}. 
Hence, this method could provide a non-destructive measurement of o-Ps annihilation in ground state with a dynamic range of about 9.5 bits inside a narrow-spaced, cryogenic, evacuated and magnetic environment. An on-leading development of this method consists of tracking the decay of laser-excited Rydberg-Ps atoms over time, which have a lifetime of the order of microseconds.

\section*{ACKNOWLEDGMENTS}
The authors wish to thank the \aegis{} collaboration at CERN for permitting the use of the positron apparatus. 
This work was supported by: Istituto Nazionale di Fisica Nucleare (INFN Italy); the CERN Fellowship program and the CERN Doctoral student program; European’s Union Horizon 2020 research and innovation program under the Marie Sklodowska-Curie grant agreements No. 754496, FELLINI.

\bibliographystyle{apsrev}
\bibliography{bib}

\end{document}